\documentclass[useAMS,usenatbib]{mn2e}
\usepackage{psfig, epsf, epsfig}
\title[The Magellanic Squall]{The Magellanic Squall:
Gas Replenishment from the Small to  Large Magellanic Cloud}
\author[K. Bekki
and M. Chiba]{Kenji Bekki${}^1$\thanks{E-mail:
bekki@phys.unsw.edu.au} and
Masashi Chiba${}^2$ \\
       ${}^1$School of Physics, University of New South Wales,
              Sydney 2052, NSW, Australia\\
      ${}^2$Astronomical Institute, Tohoku University, Sendai,
980-8578, Japan}

\begin{document}

\date{Accepted, Received 2005 February 20; in original form }

\pagerange{\pageref{firstpage}--\pageref{lastpage}} \pubyear{2005}

\maketitle

\label{firstpage}

\begin{abstract}

We first show that a large amount of metal-poor gas is stripped from
the Small Magellanic Cloud (SMC) and fallen into the Large Magellanic
Cloud (LMC) during the tidal interaction between the SMC, the LMC, and
the Galaxy over the last 2 Gyrs. We propose that this metal-poor gas
can closely be associated with the origin of LMC's young and
intermediate-age stars and star clusters with distinctively
low-metallicities with [Fe/H] $< -0.6$.
We numerically investigate whether gas initially in the outer
part of the SMC's gas disk can be stripped  during the LMC-SMC-Galaxy
interaction
and consequently can  pass through
the central region ($R<7.5$ kpc) of the LMC.
We find that about 0.7 \% and 18 \%  of the SMC's gas
can pass through the central region of the LMC about 1.3 Gyr ago
and 0.2 Gyr ago, respectively.
The possible mean metallicity of the replenished gas from the
SMC to LMC is about [Fe/H] = -0.9 to -1.0 for the two
interacting phases.
These results imply that the LMC can temporarily replenish gas supplies 
through the sporadic accretion and infall   of metal-poor gas from the SMC. 
These furthermore imply that if these gas from the SMC can collide with
gas in the LMC to form new stars in the LMC, the metallicities of
the stars can be significantly lower than those
of  stars formed from
gas initially within the LMC.
%We discuss these results in terms of the origin of distinctively
% low metallicities
%of an intermediate-age globular cluster (NGC 1718),  very young star clusters
%(e.g.  NGC 1984), and early-type stars in the inter-Cloud region close to
%the LMC.

\end{abstract}


\section{Introduction}

Tidal interaction between the LMC, the SMC, and the Galaxy
have long been considered to play vital roles
not only in dynamical and chemical evolution of the
Magellanic Clouds (MCs) but also in the formation 
of the Magellanic stream (MS) and bridge (MB)
around the Galaxy
(e.g., Westerland 1999;
Murai \& Fujimoto 1981; 
Bekki \& Chiba 2005, B05).
Although previous theoretical and numerical studies
on the LMC-SMC-Galaxy tidal interaction
discussed extensively
the origin of dynamical properties of the MB
(e.g., Gardiner \& Noguchi 1995, G96),
they have not yet investigated so extensively
the long-term formation histories of field stars and star clusters (SCs)
in the MCs.
Therefore, long-standing and remarkable problems related to
the interplay
between the LMC-SMC-Galaxy interaction
and the formation histories of stars and SCs 
remain unsolved (See Bekki et al. 2004a, b for the first
attempts to challenge  these problems).

One of intriguing and unexplained  observations on SCs in the LMC
is that an intermediate-age SC (NGC 1718)
with the estimated age of $\sim 2$ Gyr
has a distinctively low metallicity of [Fe/H]$ = -0.8$
among intermediate-age SCs 
(Geisler et al. 2003, G03;  Grocholski et al. 2006, G06).
Santos \& Piatti (2004, S04) 
investigated integrated spectrophotometric properties of
old and young SCs and found that several young SCs with ages
less than 200 Myr have metallicities smaller than $-0.6$.
Three examples of these low-metallicity objects
including Rolleston et al. (1999, R99)  are listed in
the Table 1.
Given the fact that  the stellar metallicity of the present LMC 
is about $-0.3$
in [Fe/H] (e.g., van den Bergh 2000, v00; Cole et al. 2005),
the above examples of low-metallicity, young SCs are intriguing objects.
No theoretical attempts however  have been made to understand the origin
of these intriguing objects in the LMC.

The purpose of this Letter  is to show, for the first time,
that the observed 
distinctively low metallicities
in intermediate-age and young SCs in the LMC
can be  possible evidences for accretion and infall of
low-metallicity gas onto the LMC from the SMC.
Based on dynamical simulations of the LMC-SMC-Galaxy interaction
for the last 2.5 Gyr,  we investigate whether 
gas stripped from the SMC  as a result of the tidal 
interaction can pass through the central
region of the LMC and consequently can play a role in the star
formation history of the LMC. 
Based on the results of the simulations,
we discuss how 
the sporadic accretion/infall of metal-poor gas onto the LMC from the SMC
(referred to as ``the Magellanic squall'') can control 
recent star formation activities of the LMC.

\begin{table}
\centering
\begin{minipage}{85mm}
\caption{Examples of distinctively metal-poor stars and SCs for 
the LMC and the inter-Cloud region  close to the LMC.}
\begin{tabular}{cccc}
Object name & 
Age (Myr)  & 
[Fe/H] & 
Reference   \\
NGC 1718 & 2Gyr  & $-0.80\pm0.03$  &  G06\\ 
NGC 1984 & 4Myr  & $-0.90\pm0.40$  &  S04\\ 
DGIK 975 & 41Myr  & 
{$-1.06\pm0.12$
\footnote{Mean metallicity for C, N, Ma, and Si with respect to 
the solar abundances.}}
 &  R99 \\ 
\end{tabular}
\end{minipage}
\end{table}

\section{Model}

We adopt  numerical methods and techniques of the simulations
on the evolution of the MCs
used in our previous papers (B05):
we first determine the most
plausible and realistic orbits of the MCs 
by using `` the backward integration
scheme'' (for orbital evolution  of the MCs) by Murai \&  Fujimoto (1980)
for the last 2.5 Gyr and then investigate the
evolution of the MCs using GRAPE
systems (Sugimoto et al.1990). 
The total masses of the LMC  ($M_{\rm LMC}$) and the SMC ($M_{\rm SMC}$)
are set to be $2.0 \times 10^{10} {\rm M}_{\odot}$
and $3.0 \times 10^{9} {\rm M}_{\odot}$,
respectively,  in all models.
The SMC is represented by a fully self-consistent dynamical
model with the total particle number of 200000 wheres
the LMC is represented by a point mass.
Since we focus on the mass-transfer from the SMC to the LMC,
this rather idealized way of representing the LMC 
is not unreasonable. The gravitational softening length
is fixed at 0.1 kpc for all models.

The orbital evolution of the MCs 
depends strongly on their masses and initial
velocities  for given initial locations of the MCs (B05).
Therefore we first run a large number of 
{\it collisionless} models and thereby find 
models that can successfully reproduce both the 
MS and its  leading arm features.
For those models, we investigate time evolution
of the SMC's gas particles that can finally infall onto
the LMC. 
We use the same coordinate system  $(X,Y,Z)$
(in units of kpc) as those used
in B05.
The adopted current positions
are $(-1.0,-40.8,-26.8)$  for the LMC
and $(13.6,-34.3,-39.8)$ for the SMC and
the adopted
current  Galactocentric radial velocity of the LMC (SMC)
is 80 (7) km s$^{-1}$.
Current velocities of the LMC
and the SMC in the
Galactic ($U$, $V$, $W$) coordinate
are assumed to be
(-5,-225,194) and  (40,-185,171) in units of km s$^{-1}$,
respectively. Although the adopted orbits are not
so consistent with the observationally suggested proper motion of the MCs 
by Kallivayalil et al. (2006), we think that the inconsistency
may well step from uncertainties
of the observational interpretations.

\begin{figure}
\psfig{file=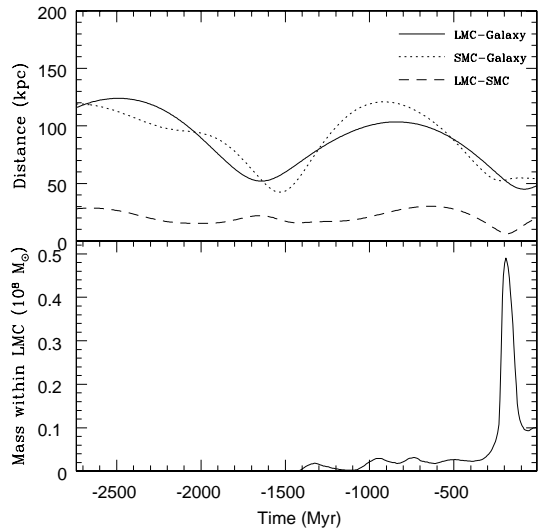,width=7.cm} 
\caption{
Time evolution of the distances between the LMC and the Galaxy
(solid), the SMC and the Galaxy (dotted), and the LMC and the SMC
(dashed), for the last 2.5 Gyrs (upper)
and time evolution  of the total mass of gas particles that
are tidally stripped from the SMC and located within
the central 7.5 kpc of the LMC (lower).
}
\label{Figure. 1}
\end{figure}

\begin{table*}
\centering
\begin{minipage}{185mm}
\caption{Model parameters for the SMC and a brief summary of the results.}
\begin{tabular}{cccccccccl}
Model number
& {Morphology 
\footnote{dE and dI denote dwarfs with spherical
stellar distributions and irregulars with disky stellar 
distributions, respectively.}}
& {$f_{\rm b}$
\footnote{Baryonic mass fraction. $f_{\rm b}$=0.18 means that the 
particle masses for dark matter, stars, and gas 
are  $ 4.9 \times 10^4 {\rm M}_{\odot}$,
$ 5.4 \times 10^3 {\rm M}_{\odot}$,
and $ 2.7 \times 10^3 {\rm M}_{\odot}$, respectively.
}} 
& {$f_{\rm g}$
\footnote{Gas mass fraction. $f_{\rm g}$=1.0 means that the gas
mass is equal to the stellar mass ($= 2.7 \times 10^8$)
in the Model 1.
}}
& { $r_{\rm g}$ 
\footnote{The size ratio of gas disk to stellar one. }}
& $\theta$ (degrees)
& $\phi$ (degrees)
& {${\rm [Fe/H]}_{\rm acc}$
\footnote{Mean metallicity of SMC's gas particles that  passed
through the central 7.5 kpc of the LMC for the last 2.5 Gyr.}}
& {$f_{\rm acc}$
\footnote{Mass fraction  of all gas particles (of the SMC)  that have  passed
through the central 7.5 kpc of the LMC for the last 2.5 Gyr. }} 
& Comment \\
1 & dE &  0.18 & 1.0 & 4.0 & -30 & 210 & -0.79 & 0.34 &  the standard model \\ 
2 & dE &  0.18 & 1.0 & 4.0 & -45 & 210 & -0.79 & 0.42 &   \\ 
3 & dE &  0.18 & 1.0 & 4.0 & -30 & 230 & -0.78 & 0.31 &   \\ 
4 & dE &  0.31 & 3.0 & 4.0 & -30 & 210 & -0.82 & 0.47 & higher gas fraction  \\ 
5 & dE &  0.18 & 1.0 & 2.0 & -30 & 210 & -0.71 & 0.65 & compact gas disk   \\ 
6 & dI &  0.18 & 1.0 & 4.0 & -30 & 210 & -0.79 & 0.34 &    \\ 
\end{tabular}
\end{minipage}
\end{table*}

Recent observations on stellar kinematics of old stars
in the SMC have suggested  that the SMC is {\it not} a dwarf
irregular with a strongly rotating stellar disk
but a dwarf spheroidal/elliptical with little rotation
(Harris \& Zaritsky 2006).  The SMC however has been modeled
as a low-luminosity disk system in
previous simulations (G96).
Considering the above observations,
we model the SMC's stellar
component  either as
a dwarf elliptical (dE) with a spherical shape
and  no rotation or as a dwarf irregular (dI)
with a disky shape and rotation
in the present study.
The SMC's stellar (gaseous) component with the size of 
$R_{\rm s}$ ($R_{\rm g}$)
and the mass of $M_{\rm s}$ ($M_{\rm g}$)
is embedded by a massive dark matter halo with
the total mass of $M_{\rm dm}$ set to be roughly equal to $9M_{\rm s}$
and  the ``universal'' density distribution
(Navarro, Frenk \& White 1996).
The projected density profile of the stellar component
has an exponential profile 
with the  scale length of $0.2 R_{\rm s}$ 
for the dE and the dI models.
$R_{\rm s}$ is fixed at 1.88 kpc so that 
{\rm almost no stellar streams can be formed along
the MS and the MB}.

Many dwarfs are observed to have extended HI gas disks
(e.g., NGC 6822; de Blok \& Walater 2003).
The SMC is therefore assumed to have an outer gas disk with
an uniform radial distribution,
$M_{\rm g}/M_{\rm s}$ ($=f_{\rm g}$), and 
$R_{\rm g}/R_{\rm s}$  ($=r_{\rm g}$)
being key parameters that determine  the dynamical evolution
of the gas.
The rotating
``gas disk' is represented by {\it collisionless particles}
in the present simulations, firstly because we
intend to understand purely tidal effects of
the LMC-SMC-Galaxy interaction on the SMC's evolution
and secondly because we compare the present results 
with previous ones by  G96 
and Connors et al. (2006) for which the ``gas'' was
represented by collisionless particles.
Although we investigate models with different $f_{\rm g}$
and $r_{\rm g}$,
we show the  results of the models with $f_{\rm g}=1$ and 3 and
$r_{\rm g}=2$ and 4 for which the Magellanic
stream with a gas mass of $\sim 10^8 {\rm M}_{\odot}$
can be reproduced reasonably well.
The baryonic mass fraction 
($f_{\rm b}=(M_{\rm s}+M_{\rm g})/M_{\rm SMC}$)
thus changes according to the adopted $f_{\rm g}$.
Owing to the adopted $r_{\rm g}=2$ and 4,
a very little amount of stars in the SMC can be transferred into
the LMC for the last 2.5 Gyr.

The initial spin of the SMC's  gas
disk in a model is specified by two angles,
$\theta$ and $\phi$, where
$\theta$ is the angle between the $Z$-axis and the vector of
the angular momentum of the disk and
$\phi$ is the azimuthal angle measured from $X$-axis to
the projection of the angular momentum vector
of the  disk onto the $X-Y$ plane.
Although these $\theta$ and $\phi$ are also considered to be free
parameters,
models with limited ranges of these parameters
can reproduce the MS and the MB (e.g., Connors et al. 2006).
The gas disk is assumed to have a {\it negative} metallicity gradient
as the stellar components has (e.g., Piatti et al. 2007).
The gradient represented by ${\rm [Fe/H]}_{\rm g}(R)$ 
(dex kpc$^{-1}$) is given as;
\begin{equation}
{\rm [Fe/H]}_{\rm g}(R)= \alpha \times R + \beta,
\end{equation}
where $R$ (in units of kpc) is the distance from the center of the SMC,
$ \alpha =-0.05$,  and $ \beta =-0.6$.
These values of  $ \alpha$ and $\beta$ are chosen
such that (i) the metallicity of the {\it central} region of
the SMC can be consistent with the observed one 
(${\rm [Fe/H]} \sim -0.6$; v00) and (ii)
the slope is well within the observed range of  $ \alpha$
for very late-type, gas-rich  galaxies (Zaritsky et al. 1994). 
If we adopt a stellar gradient (i.e., smaller $ \alpha$)
in a model,
gas particles stripped from the SMC show a smaller mean
metallicity.

We investigate (i) the time ($t_{\rm acc}$)
when gas particles 
stripped from the SMC  pass through the LMC's central 7.5 kpc
(corresponding to the disk size with the scale length of 1.5 kpc,
v00)
and (ii) the metallicities ([Fe/H]) of
the particles' for models with different
morphological types (dE or dI),  $f_{\rm b}$,
$f_{\rm g}$, $r_{\rm g}$, $\theta$, and $\phi$
in the SMC.
Such stripped SMC's particles are referred to as ``accreted particles''
in the present study just for convenience.
We also examine the mean metallicity
and the mass fraction  of the ``accreted particles''
(${\rm [Fe/H]}_{\rm acc}$ and $f_{\rm acc}$, respectively)
in each of the six models
for which values of  model parameters are shown in the Table 2. 
The present simulations
with no gas dynamics,
no  star formation, and a point-mass particle for the LMC
can not precisely predict how much
fraction of the ``accreted  particles''
can be really accreted onto the LMC's gas disk 
and consequently used for star formation.
We however believe that the present models enables us to 
grasp essential ingredients of gas transfer between
the MCs for the last few Gyrs.
We mainly show the results
for  the ``standard model'' (i.e., Model 1)
which shows typical behaviors of gas stripping in the SMC.
In the followings, the time $T$ is measured with respect
to the present-time ($T=0$): for example, $T=-1.5$ Gyr
means 1.5 Gyr ago in the present study.

\begin{figure}
\psfig{file=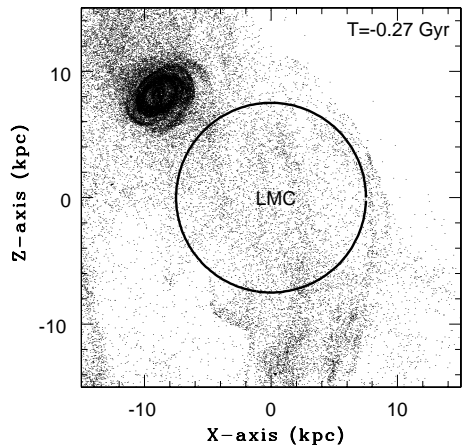,width=6.cm} 
\caption{
The distribution of gas from the SMC with respect
to the LMC's center at $T=-0.27$ Gyr. The circle
represents the disk radius of the LMC.
}
\label{Figure. 2} 
\end{figure}

\begin{figure}
\psfig{file=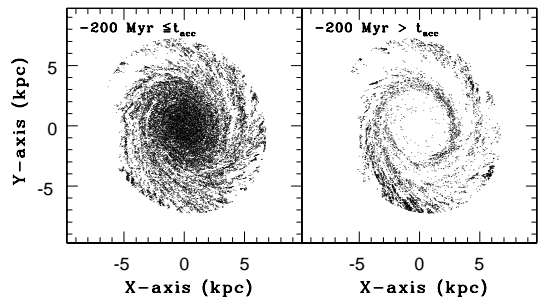,width=7.cm} 
\caption{
The initial locations (with respect to the SMC's center)
of gas particles that
are stripped from the SMC and then  pass through 
the central 7.5 kpc of the LMC
before (right) and after (left)
$T=-200$ Myr in the standard model.
$t_{\rm acc}$ denotes the time when
a particle passes through the central region of the LMC
{\it last time}.
}
\label{Figure. 3} 
\end{figure}

\begin{figure}
\psfig{file=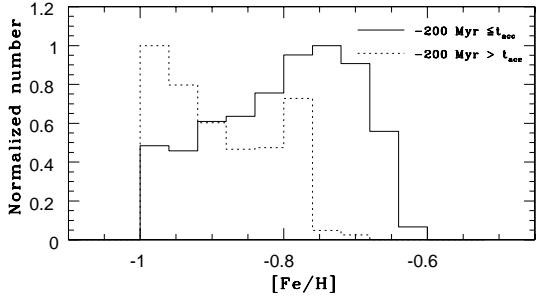,width=7.cm} 
\caption{
The metallicity distribution (MD) of gas particles
with $-200$ Myr $ \le t_{\rm acc}$  (solid)
and $t_{\rm acc} < -200$ Myr (dotted)
in the standard model. This MD depends strongly
on initial metallicity gradients.
}
\label{Figure. 4} 
\end{figure}

\begin{figure}
\psfig{file=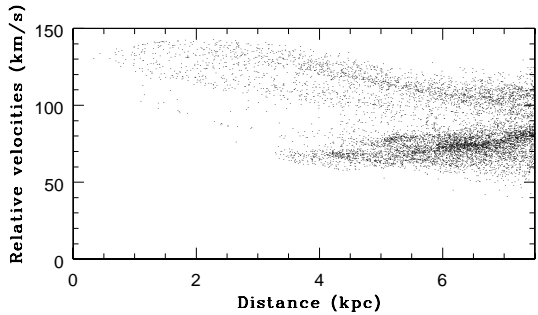,width=7.cm} 
\caption{
Plots of one-dimensional relative velocities  ($V_{\rm rel}$)
of the SMC's gas particles
as a function of their distances from the center of the LMC
at $T=-200$ Myr in the standard model.
Here $V_{\rm rel}$ for each particle is estimated as 
$V_{\rm rel}
={ ( ({v_{\rm lx}}^2+ {v_{\rm ly}}^2+ {v_{\rm lz}}^2)/3) }^{0.5}$,
where $v_{\rm lx}$,  $v_{\rm ly}$, and  $v_{\rm lz}$
are $X-$, $Y-$, and $Z-$components of the relative
velocity of the particle with respect to the LMC's velocity,
respectively.
}
\label{Figure. 5} 
\end{figure}

\section{Results}

Fig.1 shows, for the standard model (Model 1), the time evolution of
the total gas mass (stripped from the SMC) which reaches and is just
located within the central 7.5 kpc of the LMC at each time step, 
$M_{\rm acc}$.
It is noted that $M_{\rm acc}$
 is not an accumulated gas mass but is changeable
with time as gas particles can pass through the LMC in the current
collisionless simulation.
It is clear that the $M_{\rm acc}$ evolution  shows  a number of peaks 
with the first peak about $T=-1.3$ Gyr 
($M_{\rm acc}=1.8 \times 10^6 {\rm M}_{\odot}$),
just after the first pericenter passage of the SMC with
respect to the Galaxy in the 2.5 Gyr evolution. 
The highest peak is seen at $T=-200$ Myr 
($M_{\rm acc}=4.9 \times 10^7 {\rm M}_{\odot}$),
when the LMC and the SMC interact the most strongly.
Since the  gas mass ($M_{\rm acc}$) at its peak is not negligibly small
compared with the present-day HI mass of the LMC
($7.0 \times 10^8 {\rm M}_{\odot}$; v00),
accretion and infall of the gas onto the LMC's gas disk
can increase local gas densities and consequently can possibly trigger
star formation in the LMC.
Fig. 2 demonstrates the epoch of the ``Magellanic squall'',
when the stripped gas particles 
of the SMC are falling onto  the disk  of the LMC. 

Fig. 3 shows the initial locations of the SMC's gas particles
(with respect to the SMC's center)  with $-200$ Myr $ \le t_{\rm acc}$  
and  $t_{\rm acc} < -200$ Myr,
where $t_{\rm acc}$ denotes the time when a particle passed through
the central region of the LMC last time.
The particles with  $ t_{\rm acc} < -200$ Myr are initially
located in the outer part of the SMC so that they can be stripped
from the SMC and consequently pass through the LMC earlier.
Owing to the small pericenter distance of the LMC-SMC orbital
evolution at $T=-200$ Myr, 
the SMC is strongly disturbed to lose gas particles not only  from
its outer part but from its inner one.
As a result of this,  gas initially located throughout the gas
disk of the SMC can pass through the central region of the LMC
at  $T=-200$  Myr and thus show $-200$ Myr $ \le t_{\rm acc}$.

The abovementioned  differences in the initial spatial distributions
between gas particles with  $-200$ Myr $ \le t_{\rm acc}$
and $t_{\rm acc} < -200$ Myr  
can cause  the differences in metallicity distributions
of the gas between the two populations,
because the SMC's gas disk is assumed to have a negative metallicity gradient.
Fig. 4 shows that the gas particles with $t_{\rm acc} < -200$ Myr
have a larger fraction of metal-poor gas 
with $-0.6 \le$[Fe/H]$\le -1.0$ and a mean metallicity of [Fe/H]$=-0.86$.
The particles with  $t_{\rm acc} \approx  -1.3$ Gyr has a mean
metallicity of [Fe/H]$=-0.95$, because they are initially located
in the outermost part of the SMC's gas disk.
Fig. 4 also shows that
the gas particles with  $-200$ Myr $ \le t_{\rm acc}$
have a peak around  [Fe/H]$=-0.7$ with
a mean metallicity of [Fe/H]$=-0.77$. 
The particles with  $t_{\rm acc} \approx  -200$ Myr has a mean
metallicity of [Fe/H]$=-0.86$.
These results clearly suggest that the LMC can replenish gas supplies
through accretion and infall of {\it metal-poor gas from the SMC}
onto the LMC's disk.
It should be stressed here that the metallicities of
accreted gas from the SMC at $t_{\rm acc} \approx  -200$ Myr
can be appreciably higher than the above, if we consider
chemical evolution of the SMC due to star formation
for the last 2.5 Gyr.

Fig. 5 shows that relative velocities ($V_{\rm rel}$)
of the SMC's gas particles within the central 7.5 kpc of the LMC
with respect to the LMC velocity range from 40 to 150 km s$^{-1}$ at
$T=-200$ Myr.
This result indicates that if the particles can infall onto 
the LMC's disk, they can give strong dynamical impact
on the HI gas of the LMC and possibly cause shock energy dissipation
owing to $V_{\rm rel}$ much higher than the sound velocities
of cold gas. 
Previous numerical simulations showed that cloud-cloud collisions
with moderately high relative velocities ($V_{\rm rel}=10-60$km s$^{-1}$)
can trigger the formation of SCs (Bekki et al. 2004a).
Therefore the above result implies that some fraction of the 
particles passing through the LMC's central region can be responsible
for the formation of new SCs in the LMC.

The parameter dependences of 
${\rm [Fe/H]}_{\rm acc}$ and $f_{\rm acc}$
are briefly summarized as follows.
Firstly ${\rm [Fe/H]}_{\rm acc}$ and $f_{\rm acc}$
do not depend so strongly on baryonic fractions,
gas mass fractions, and orbital configurations (See the Table 2):
${\rm [Fe/H]}_{\rm acc}$ ($f_{\rm acc}$)
ranges from $-0.79$ (0.34) to $-0.82$ (0.47) for a fixed size ratio
of $r_{\rm g}$ ($=R_{\rm g}/R_{\rm s}$).
Secondly, ${\rm [Fe/H]}_{\rm acc}$ and $f_{\rm acc}$ are 
{\it both larger}
in the model with smaller $r_{\rm g}$ (Model 6) for
which a smaller amount 
of gas particles can be tidally stripped from the SMC.
The reason for
the larger $f_{\rm acc}$  is that
a significantly larger fraction of particles once stripped
from the SMC can pass through the LMC in Model 6.
Thirdly, the morphological type of
the SMC in the present study is not important
for ${\rm [Fe/H]}_{\rm acc}$ and $f_{\rm acc}$. 
Given the fact that only 0.2\% of gas can be converted into
strongly bound SCs (rather than into field stars) in
the evolution of the MCs (B05),
these results imply that
the maximum possible mass of SCs formed from SMC's gas
in the LMC is roughly $10^5 {\rm M}_{\odot}$ in the present
models. Owing to the very short time scale 
($\sim 10^6$ yr) of SC formation
from gas clouds during the tidal interaction (Bekki et al. 2004b),
the stripped SMC's gas clouds are highly likely to
be accreted onto the LMC within the dynamical time scale
of the LMC ($\sim 10^8$ yr) and then converted into SCs
within $\sim 10^6$ yr after the accretion.

\section{Discussions and conclusions}

NGC 1718 with an estimated age of $\sim 2$ Gyr 
has a low metallicity ([Fe/H]$\sim -0.8$)
about 0.3 dex smaller than those of other SCs with
similar ages 
in the LMC (e.g., G03; G06).
If the interstellar medium (ISM) of the LMC about $1-2$ Gyr ago
was very inhomogeneous in terms of chemical abundances,
some fraction of stars could be born from quite low-metallicity
gas clouds with [Fe/H]$\sim -0.8$.
The distinctively low metallicity therefore could be due to  
the abundance inhomogeneity of the ISM in the LMC about $1-2$ Gyr ago.
However,
intermediate-age SCs {\it other than} NGC 1718  have very similar
metallicities of [Fe/H]$\sim -0.48$ and  a small metallicity dispersion
of only 0.09 dex  in the LMC (G06).
The observed low-metallicity of NGC 1718 thus seems to be unlikely to
be due to the abundance inhomogeneity of the ISM. 
We suggest that the origin of the NGC 1718
can be closely associated with
the Magellanic squall about $1-2$ Gyr ago. 
Since gaseous abundance patterns (e.g., [Mg/Fe]) of the SMC
about a few Gyr ago might well be very different from 
those of the LMC,  NGC 1718 could have abundance patterns
quite different from those of other GCs.
It should be here stressed that the simulated peak of the squall
($\sim 1.3$ Gyr) is not very consistent with the observed 
age of NGC 1718 ($2.0 \pm 0.4$ Gyr, G03).

S04 recently have reported that
eight  young SCs with ages less than 200 Myr 
have metallicities smaller than [Fe/H]$=-0.3$ that is
a typical stellar metallicity of the LMC (e.g., v00).
Although there could be some observational uncertainties in
age and metallicity determination based solely on integrated 
spectrophotometric properties 
of SCs (S04),
their results imply that these SCs could have been formed from
metal-poor gas in the LMC quite recently.
The present numerical results imply  that
NGC 1711, NGC 1831, NGC 1866,  and NGC 1984, all of which
are observed to  have possible metallicities smaller than [Fe/H]$=-0.6$,
can be formed as a result of the Magellanic squall.
Since the chemical abundances of the outer gas disk of the SMC
can be significantly different from those of the present LMC's gas
disk,  the detailed abundances 
(e.g., [C/Fe],  [N/Fe], and [Mg/Fe]) of the above four clusters
can be significantly different from those of other
young SCs with ``normal'' metallicities with [Fe/H]$=  -0.3 - -0.5$
in the LMC.
The observed young, metal-poor
stars ([Fe/H]$\sim -1.0$) in the inter-Cloud region
close to the LMC (R99) will be equally explained by the gas-transfer
between the MCs (see also Bekki \& Chiba 2007).

The present study has first pointed out that
the Magellanic squall can also play a role
in the  relatively recent star formation history of the LMC.
Sporadic infall of metal-poor gas like the Magellanic squall
might well be also important for recent star formation histories
in pairs of interacting galaxies.
Previous hydrodynamical simulations showed that
high-velocity collisions of HI gas onto a galactic disk
can create HI holes and shells (e.g., Tenorio-Tagle et al. 1986).
The Magellanic squall, which inevitably can cause high-velocity
impact of the gas clouds stripped from the SMC on the LMC,
can  thus be responsible for {\it some} of the observed HI holes
in the LMC (e.g.,  Staveley-Smith et al. 2003).
We plan to investigate how collisions between low-metallicity
gas clouds from the SMC and those initially in the LMC
trigger the formation of stars and SCs in the LMC's disk
based on more sophisticated, high-resolution hydrodynamical simulations
with pc-scale  star formation processes.
Our future studies thus will enable us to understand more deeply
how the Magellanic squall influences pc-scale star formation
processes in the LMC.

\section{Acknowledgment}
We are  grateful to the  referee, Daisuke Kawata,
for his valuable comments,
which contribute to improve the present paper.
K.B. acknowledges the Large Australian Research Council (ARC).
Numerical computations reported here were carried out on GRAPE
system kindly made available by the Astronomical Data Analysis
Center (ADAC) of the National Astronomical Observatory of Japan.

\end{document}